\documentclass{ptephy_v1}%%%%%% to generate preprint number with ptep logo

\preprintnumber{XXXX-XXXX} %%% %%% Insert preprint number here

\usepackage{graphicx}
\usepackage{natbib}
\usepackage{times}
\usepackage{amsmath,mathrsfs,mathtools,color,wasysym,dsfont,here}
\usepackage{amsthm}
\usepackage{physics}
\usepackage{url} %It provides better support for handling and breaking URLs.

\begin{document}

\title{Quantum Ising chain with boundary dephasing}

%%%% To generate auto affiliation numbers please use \author{}\affil{} command

\author{Naoyuki Shibata}
\affil{Department of Physics, Graduate School of Science,\\ 
	the University of Tokyo, 7-3-1 Hongo, Tokyo 113-0033, Japan}

\author[1, 2, 3]{Hosho Katsura}
\affil{Institute for Physics of Intelligence, The University of Tokyo, 7-3-1 Hongo, Tokyo 113-0033, Japan}
\affil{Trans-scale Quantum Science Institute, The University of Tokyo, 7-3-1 Hongo, Tokyo 113-0033, Japan}

%%% To include the collaborator name... Please use the command "\collaborator"
%%% For example: \collaborator{ATLAS Collaboration}

\begin{abstract}%
    We study the quantum Ising chain with boundary dephasing. By doubling the Hilbert space, the model is mapped to the Su-Schrieffer-Heeger model with imaginary chemical potential at the edges. We show analytically and numerically that the Liouvillian gap, i.e., the inverse relaxation time of the model, scales with the system size $ N $ as $ N^{-3} $.
\end{abstract}

\subjectindex{xxxx, xxx}

\maketitle

\section{Introduction}
    Understanding the dynamics of open quantum systems is of fundamental importance in a variety of fields including quantum information~\cite{Kraus2008,Kastoryano2011}, quantum computing~\cite{Verstraete2009}, and condensed matter physics~\cite{Diehl2008,Diehl2010,Diehl2011,Bardyn2013}. The Lindblad equation~\cite{Gorini1976, Lindblad1976} is widely used to describe such open quantum systems under the assumption of Markovian and completely positive and trace preserving (CPTP) dynamics. Although in the past this quantum master equation had been mainly applied to few-particle systems in, e.g., quantum optics, the Lindblad equations for many-particle systems have recently attracted much attention due to the recent advances in quantum engineering. 
    
    Exactly solvable models play an important role in understanding the intricate physics of quantum many-body systems. Recently, several dissipative but still exactly solvable models have been studied. Roughly speaking, solvability/integrability of these models mainly relies on free-fermion or free-boson techniques~\cite{Prosen2008,Prosen2008a,Prosen2010,Guo2017,Shibata2019} or the Bethe (or similar tensor network) ansatz~\cite{Prosen2011,Prosen2013,Medvedyeva2016,Rowlands2018,Shibata2019a,Nakagawa2020,Buca2020,Ziolkowska2020}. In this paper, we show a pedagogical example of a dissipative solvable model categorized in the former case. We consider the quantum Ising chain with the open boundary conditions in the presence of dephasing at both edges. By vectorizing the density matrix, the model can be mapped to a non-Hermitian tight-binding model, i.e., the Su-Schrieffer-Heeger (SSH) model~\cite{Su1979} with imaginary chemical potential at the edges, and thus solved exactly. This mapping allows for a detailed analysis of the dynamics. In particular, we study how the Liouvillian gap $ g $ which usually corresponds to the inverse of the relaxation time scales with the system size $ N $ both analytically and numerically. The analysis reveals that the gap $ g $ is inversely proportional to the cube of $ N $ for large $ N $.
    
    The rest of the paper is organized as follows. In Sec. \ref{sec:model}, we define our model and show how it is mapped to a free-fermion model. In Sec. \ref{sec:Liouvillian_gap}, we investigate the spectrum of the Liouvillian, the generator of time evolution. More precisely, in Sec. \ref{subsec:general_review}, we review the general concepts of the Liouvillian spectrum. In Sec. \ref{subsec:spectrum_of_our_model}, we study the Liouvillian spectrum of our model both analytically and numerically. In Sec. \ref{subsec:plane_wave}, we further investigate the Liouvillian spectrum in the case where the parameters are fine-tuned using the plane wave ansatz. We present our conclusions in Sec. \ref{sec:conclusion}. In Appendix \ref{app:qunatum_compass}, we present another model with the boundary dephasing which reduces to the same non-Hermitian free-fermion model.
    
\section{Model}\label{sec:model}
    We consider the time evolution of the density matrix $ \rho $ governed by the Lindblad equation (here and in what follows we set $ \hbar=1 $)
	\begin{align}
		\dv{\rho}{t}=\mathcal{L}[\rho]\coloneqq-\mathrm{i}[H,\rho]+\sum_{i=\mathrm{L},\mathrm{R}}\qty(L_i\rho L_i^\dagger-\dfrac{1}{2}\qty{L_i^\dagger L_i,\rho}).\label{eq:Lindblad}
	\end{align}
	A non-dissipative part of the model is a quantum Ising chain with open boundary conditions described by the Hamiltonian
	\begin{align}
		H=H_\mathrm{QI}&=-h\sum_{i=1}^{N}\sigma_i^x -J\sum_{i=1}^{N-1} \sigma_i^z\sigma_{i+1}^z,\label{eq:hamQI}
	\end{align}
	where $ N $ is the number of sites and  $ \sigma_i^\alpha $ $ (i=1,\dots, N,\; \alpha=x,y,z) $ are Pauli matrices at site $ i $. The Lindblad operators acting on the left and right edges are dephasing written as
	\begin{align}
	\begin{split}
	    L_\mathrm{L}&=\sqrt{\gamma_\mathrm{L}}\sigma_1^z,\\ L_\mathrm{R}&=\sqrt{\gamma_\mathrm{R}}\sigma_N^z,
	\end{split}\label{eq:boundary_dis}
	\end{align}
	where $ \gamma_\mathrm{L}, \gamma_\mathrm{R}\ge 0 $ are dephasing strengths. Without loss of generality, we can assume parameters $ h, J \ge 0 $, as the other cases can be obtained by an appropriate unitary transformation. It is well known that the original quantum Ising model exhibits quantum phase transition at $ h=J $ and the ground state is \textit{ordered} (\textit{paramagnetic}) when $ h < J $ ( $ h > J) $~\cite{Sachdev1999}.  The generator $ \mathcal{L} $ of the dynamics is called a \textit{Liouvillian} or a \textit{Lindbladian}. The Liouvillian is a superoperator, i.e., an operator acting in the space of operators. Since the Liouvillian itself is a linear map, it can be seen as a matrix if we identify each linear operator on the Hilbert space with a vector. As a result, the $ 2^N\times 2^N $ density matrix $ \rho $ is vectorized as a $ 2^{2N} $-dimensional vector. Through this mapping, the Liouvillian $ \mathcal{L} $ (times $ \mathrm{i} $) is identified as a non-Hermitian Hamiltonian as~\cite{Minganti2018,Shibata2019,Shibata2019a}
	\begin{align}
		\mathrm{i}\mathcal{L}&\cong\mathcal{H}\coloneqq H\otimes\mathds{1}-\mathds{1}\otimes H^\mathrm{T}+\mathrm{i}\sum_{i}\qty(L_i\otimes L_i^\ast-\dfrac{1}{2}L_i^\dagger L_i\otimes\mathds{1}-\dfrac{1}{2}\mathds{1}\otimes L_i^\mathrm{T}L_i^\ast),\label{eq:Lindbladian_mapped_to_non-Hermitian_Hamiltonian}
	\end{align}
	where the Hilbert space of the RHS is the ``$ \mathrm{Ket}\otimes\mathrm{Bra} $ space''. In our model, the corresponding Hamiltonian $ \mathcal{H}=\mathcal{H}_\mathrm{QI} $ reads
	\begin{align}
		\mathcal{H}_\mathrm{QI}=& -h\sum_{i=1}^{N}\sigma_i^x - J\sum_{i=1}^{N-1} \sigma_i^z\sigma_{i+1}^z + h\sum_{i=1}^{N}\tau_i^x + J\sum_{i=1}^{N-1}\tau_i^z\tau_{i+1}^z 
		\nonumber \\
		& + \mathrm{i}\gamma_\mathrm{L}\,\sigma_1^z\tau_1^z+\mathrm{i}\gamma_\mathrm{R}\,\sigma_N^z\tau_N^z-\mathrm{i}\,(\gamma_\mathrm{L}+\gamma_\mathrm{R}),
	\end{align}
	where $ \tau_i^\alpha $ ($ \alpha=x,y,z $) are the Pauli matrices for the $ i $th Bra site. This model has a conserved charge, that is, the \textit{parity operator}
	\begin{align}
		Q\coloneqq \qty(\prod_{j=1}^{N}\sigma_j^x)\qty(\prod_{j=1}^{N}\tau_j^x)
	\end{align}
	with eigenvalues $ \pm 1 $.
	With the Jordan-Wigner transformation
	\begin{gather}
		a_i=(-1)^i\qty(\prod_{j=1}^{i-1}\sigma_j^x)\sigma_i^z,\quad b_i=(-1)^i\qty(\prod_{j=1}^{i-1}\sigma_j^x)\sigma_i^y,\notag\\
		\bar{a}_{N+1-i}=(-1)^{N+1-i}\qty(\prod_{j=1}^{N}\sigma_j^x)\qty(\prod_{j=1}^{i-1}\tau_{N+1-j}^x)\tau_{N+1-i}^z,\notag\\
		\bar{b}_{N+1-i}=(-1)^{N-i}\qty(\prod_{j=1}^{N}\sigma_j^x)\qty(\prod_{j=1}^{i-1}\tau_{N+1-j}^x)\tau_{N+1-i}^y ,
	\end{gather}
	the model is mapped to a free Majorana fermion model
	\begin{align}
		\mathcal{H}_\mathrm{QI}=&-h\sum_{i=1}^{N} \qty(\mathrm{i}a_i b_i+\mathrm{i}\bar{a}_i\bar{b}_i) + J\sum_{i=1}^{N-1}\qty(\mathrm{i}b_i a_{i+1}+\mathrm{i}\bar{b}_{i+1}\bar{a}_i)
		\nonumber \\
		&-\gamma_\mathrm{L}\, Q\bar{b}_1 a_1 -\gamma_\mathrm{R}\,b_N \bar{a}_N - \mathrm{i}\,(\gamma_\mathrm{L} + \gamma_\mathrm{R}),
	\end{align}
	where $ a_i, b_i, \bar{a}_i $ and $ \bar{b}_i $ are Majorana operators. We then define annihilation/creation operators $ c_i/c_i^\dagger $ of complex (Dirac) fermions
	\begin{gather}
		a_i=c_{2i-1}+c_{2i-1}^\dagger,\quad \bar{b}_i=\dfrac{1}{\mathrm{i}}(c_{2i-1}-c_{2i-1}^\dagger),\\
		\bar{a}_i=c_{2i}+c_{2i}^\dagger,\quad b_i=\dfrac{1}{\mathrm{i}}(c_{2i}-c_{2i}^\dagger)
	\end{gather}
	to rewrite the model as
	\begin{align}
	    \begin{split}
		\mathcal{H}_\mathrm{QI}&=-2h\sum_{i=1}^{N}\qty(c_{2i-1}^\dagger c_{2i}+\textrm{h.c.})-2J\sum_{i=1}^{N-1} \qty(c_{2i}^\dagger c_{2i+1}+\textrm{h.c.})\\
		&\hspace{1em}-2\mathrm{i}\gamma_\mathrm{L}Q\qty(c_1^\dagger c_1-\dfrac{1}{2}) -2\mathrm{i}\gamma_\mathrm{R}\qty(c_{2N}^\dagger c_{2N} -\dfrac{1}{2})-\mathrm{i}(\gamma_\mathrm{L}+\gamma_\mathrm{R})
		\end{split}\label{eq:SSH_with_boundary_potential}\\
		&=-2\vb{c}^\dagger \mathsf{T}_\mathrm{QI} \vb{c} +\mathrm{i}\gamma_\mathrm{L}(Q-1).\label{eq:H_QI_SSH}
	\end{align}
	Here, $ \vb{c}=(c_1, \dots, c_{2N})^\mathrm{T}, \vb*{c}^\dagger=(c_1^\dagger, \dots, c_{2N}^\dagger), $ and 
	\begin{align}
		\mathsf{T}_\mathrm{QI}\coloneqq
		\begin{pmatrix}
			\mathrm{i}\gamma_\mathrm{L}Q & h & & & &\\
			h & 0 & J & & &\\
			& J & 0 & \ddots & &\\
			& & \ddots & \ddots & \ddots &\\
			& & & \ddots & 0 & h\\
			& & & & h & \mathrm{i}\gamma_\mathrm{R}
		\end{pmatrix}\label{eq:SSH_QI}
	\end{align}
	is a $ 2N\times 2N $ tridiagonal matrix referred to as a one-particle Hamiltonian. It is nothing but a Su-Schrieffer-Heeger (SSH) model~\cite{Su1979} with imaginary chemical potential acting on the edges. One can see that the parity operator $ Q $ is rewritten as
	\begin{align}
		Q=(-1)^{\hat{F}},
	\end{align}
	where $ \hat{F}=\sum_{i=1}^{2N}c_i^\dagger c_i $ is the total number of complex fermions.
	
	Before closing this section, several remarks are in order. First, when $ \gamma_\mathrm{L}=\gamma_\mathrm{R}=0 $, i.e., without dephasing, there is a correspondence between phases of the quantum Ising model and the SSH model. That is, when $ h<J $, the ground state of the quantum Ising model is in the \textit{ordered} phase, as mentioned above, while the SSH model is in the \textit{topological} phase. Meanwhile, when $ h>J $, the ground state of the quantum Ising model is in the \textit{paramagnetic} phase, while the SSH model is in the \textit{trivial} phase.  Second, for $ Q=-1 $ sector, the model has $ \mathcal{PT} $-symmetry  when $ \gamma_\mathrm{L}=\gamma_\mathrm{R} $, i.e., it is invariant under the combination of parity (spatial reflection) and time-reversal transformation. Here, parity $ \hat{\mathcal{P}} $ and time-reversal $ \hat{\mathcal{T}} $ act as
	\begin{align}
	    \hat{\mathcal{P}}c_j\hat{\mathcal{P}}^{-1}=c_{2N-j+1},\quad \hat{\mathcal{T}}\,\mathrm{i}\,\hat{\mathcal{T}}^{-1}=-\mathrm{i}.
	\end{align}
	This particular case was already intensively studied in Ref.~\cite{Klett2017}. Third, the same one-particle Hamiltonian as Eq.~(\ref{eq:SSH_QI}) arises from another dissipative model, see Appendix \ref{app:qunatum_compass}.
	
% 	In the following, we further investigate the properties of the model described by Eq.~(\ref{eq:SSH_with_boundary_potential}).
	
\section{Liouvillian spectrum}\label{sec:Liouvillian_gap}
    \subsection{General review}\label{subsec:general_review}
\begin{figure}[!h]
    		\centering
    		\includegraphics[width=0.65\linewidth]{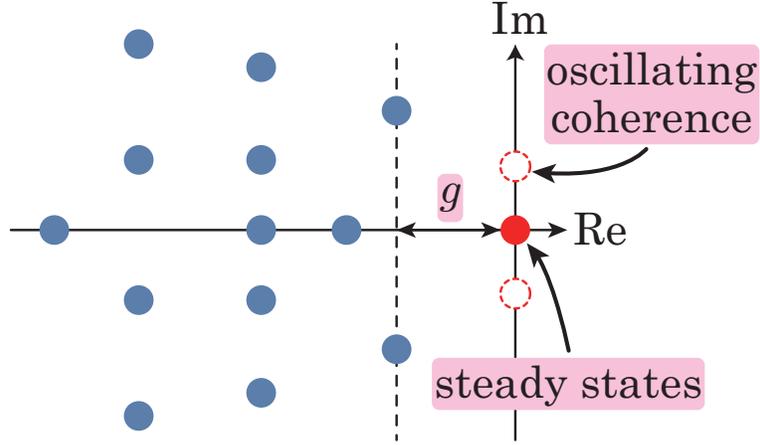}
    		\caption{A schematic of the Liouvillian spectrum. The solid red dot denotes the steady state, while red dashed circles denote oscillating coherences. Other blue dots correspond to decay modes which are vanishing as $ t\to\infty $. The Liouvillian gap $ g $ is equal to the inverse of the longest relaxation time of decay modes.}
    		\label{fig:liouvillian_spectrum}
    	\end{figure}

        In this section, we explain some important characteristics of the Liouvillian and its spectrum. Let $ \Lambda_i $ be an eigenvalue of the Liouvillian $ \mathcal{L} $. Then the following holds~\cite{Breuer2002,Rivas2011,Minganti2018}:
    	\begin{enumerate}
    		\item All $ \Lambda_i $ satisfy $ \Re(\Lambda_i)\le 0 $.
    		
    		\item If $ \mathcal{L}[\rho_i]=\Lambda_i\rho_i $, then $ \mathcal{L}\qty[\rho_i^\dagger]=\Lambda_i^\ast\rho_i^\dagger $.
    	\end{enumerate}
    	Figure \ref{fig:liouvillian_spectrum} shows a schematic example of the Liouvillian spectrum. A \textit{non-equilibrium steady state} (\textit{NESS}) is an eigenstate of the Liouvillian $ \mathcal{L} $ with eigenvalue 0: $ \mathcal{L}[\rho_\infty]=0 $. By definition, such a state is time-independent and free from decoherence. In other words, steady states have infinite lifetimes. Similarly to NESSs, states with nonzero purely imaginary eigenvalues are called \textit{oscillating coherences} \cite{Albert2014} or \textit{undamped oscillating phase relations} \cite{Baumgartner2008}. They are time-dependent but free from decay. Other states which have finite lifetimes are called \textit{decay modes}.
    			
    	The main object of our interest is a \textit{Liouvillian gap} $ g $ defined by \cite{Znidaric2015}
    	\begin{equation}
    		g\coloneqq -\max_{\substack{i\\ \Re(\Lambda_i)\ne 0}}\Re(\Lambda_i),\label{eq:def_of_Liouvillian_gap}
    	\end{equation}
    	which corresponds to the inverse of the relaxation time, i.e, the longest lifetime of the decay modes. An eigenstate of $ \mathcal{L} $ such that $ \Re(\Lambda_i)= -g $ is called the \textit{first decay mode}.
    	%Eigenstates with eigenvalue $ \Lambda_i $ which satisfies $ g=-\Re(\Lambda_i) $ is called the first decay mode.
    	
    \subsection{Spectrum of Eq.~(\ref{eq:H_QI_SSH})}\label{subsec:spectrum_of_our_model}
\begin{figure}[!h]
            \centering
            \includegraphics[width=0.65\linewidth]{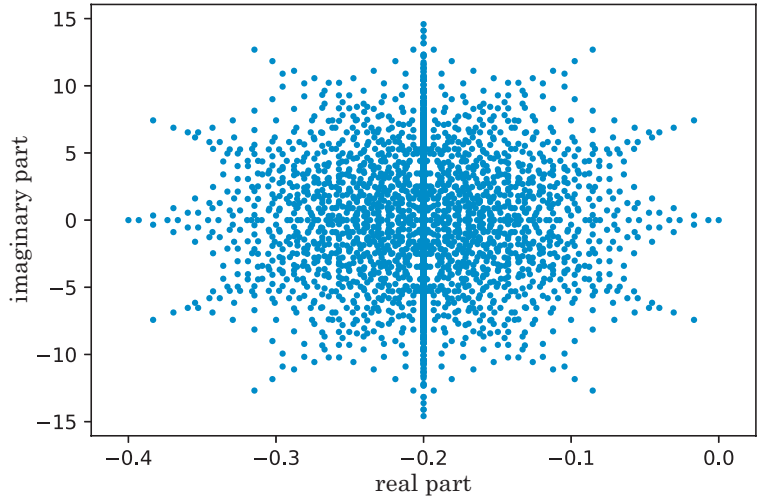}
            \caption{Liouvillian spectrum of our model with $ N=8,\; h=J=1.0,\; \gamma_\mathrm{L}=\gamma_\mathrm{R}=0.1 $.}
            \label{fig:Liouvillian_spectrum_N_6_fig}
        \end{figure}

        We shall now discuss the Liouvillian spectrum of our model. For simplicity, we fix $ \gamma_\mathrm{L}=\gamma_\mathrm{R}=\gamma $  in the following~\footnote{The case where dissipation acts only on one edge has been studied in Ref.~\cite{Vasiloiu2018}.}. We show in Fig.~\ref{fig:Liouvillian_spectrum_N_6_fig} a Liouvillian spectrum of our model obtained by exact diagonalization. One can see that the spectrum has a dihedral group ($ D_2 $) symmetry in the complex plane with the lines of symmetry $ l_\mathrm{v}=-2\gamma+\mathrm{i}\mathbb{R} $ and $ l_\mathrm{h}=\mathbb{R} $. It is because the Liouvillian has so-called \textit{Liouvillian $ \mathbb{PT} $-symmetry}~\cite{Prosen2012,Huybrechts2020}, which is easily verified by taking $ U=\prod_{i=1}^N \sigma_i^x,\, W=\mathds{1} $ in Ref.~\cite{Prosen2012a}. Furthermore, Fig.~\ref{fig:Liouvillian_spectrum_N_6_fig} suggests that there is a NESS but no oscillating coherence. In the following, we investigate the details.
        
        First, let us consider the NESS. Forgetting the above mapping to non-Hermitian Hamiltonian for the moment, we go back to the original Lindblad equation~(\ref{eq:Lindblad}). Since two Lindblad operators~(\ref{eq:boundary_dis}) are both Hermitian, the completely mixed state $ \rho_\mathrm{0}=\mathds{1}/2^N $ is a NESS\footnote{When the dimension of the Hilbert space is finite, this statement is easily verified since $ \mathcal{L}[\mathds{1}]=0 $.}, where we assume $ N $ is finite, and so is the dimension of the Hilbert space. We numerically confirmed that it is the only NESS of our model.
        
        Next, let us consider the Liouvillian gap of our model. It is clear from Eqs. (\ref{eq:def_of_Liouvillian_gap}) and (\ref{eq:Lindbladian_mapped_to_non-Hermitian_Hamiltonian}) that the Liouvillian gap corresponds to the gap between the first and the second largest \textit{imaginary} parts of the eigenvalues of $ \mathcal{H}_\mathrm{QI} $. The eigenvalue of $ \mathcal{H}_\mathrm{QI} $ which has the largest imaginary part lives in $ Q=+1 $ sector. In fact, when none of the one-particle eigenvalues of $ -\mathsf{T}_\mathrm{QI} $ is filled, one obtains many-body eigenvalue $ 0 $ of $ \mathcal{H}_\mathrm{QI} $.
        
\begin{figure}
            \centering
            \includegraphics[width=1.0\linewidth]{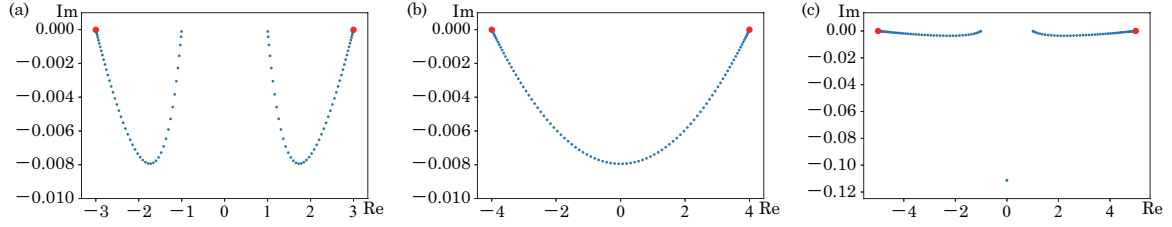}
            \caption{One-body eigenvalues of $ -2\mathsf{T}_\mathrm{QI} $ with $ 2N=100, \gamma=0.1 $, and (a) $ h=1.0,\; J=0.5 $ (trivial), (b) $ h=1.0,\; J=1.0 $ (critical), and (c) $ h=1.0, \; J=1.5 $ (topological). Imaginary parts of all eigenvalues are negative and red dots denote the eigenvalue with the smallest imaginary part in absolute value. These red dots also have the largest real part in absolute value.}
            \label{fig:SSH_L_100_fig}
        \end{figure}

\begin{figure}
            \centering
            \includegraphics[width=0.65\linewidth]{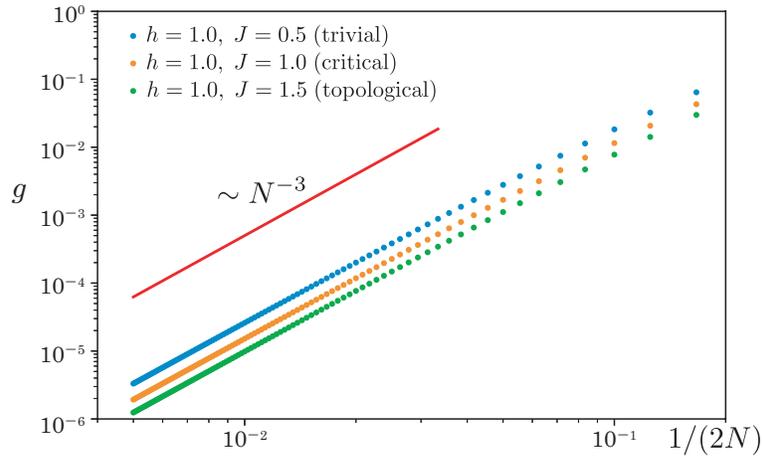}
            \caption{Numerically calculated Liouvillian gap $ g $ for small $ \gamma=0.1 $. The red line is a guide to the eye indicating the $ N^{-3} $ scaling.}
            \label{fig:Liouvillian_gap_scaling}
        \end{figure}

        We numerically confirmed that the Liouvillian gap is determined by the spectral gap in $ Q=+1 $ sector and scales as $ \sim N^{-3} $ regardless of whether $ h<J $ or $ h>J $. Thus, the Liouvillian spectrum is gapless in the thermodynamic limit. Such a cubic scaling of the Liouvillian gap is also found in several other models~\cite{Prosen2008,Prosen2008a,Znidaric2015}. Scaling of $ g $ can be explained using perturbation theory. Figure \ref{fig:SSH_L_100_fig} shows the eigenvalues of $ -2\mathsf{T}_\text{QI} $ for $ L=100 $ and $ \gamma=0.1 $ in each phase. As one can see, the eigenvalues with the largest real part in absolute value (shown by the red dots) have the smallest imaginary part in absolute value. Then, the many-body eigenvalues of $ \mathcal{H}_\mathrm{QI} $ corresponding to the first decay mode is obtained by filling only this one-body eigenvalue. Thus, we apply non-degenerate perturbation theory for the state with the maximum modulus eigenvalue. The unperturbed eigenvalue $ E_\mathrm{M} $ and the (unnormalized) eigenvector $ \vb*{x}^\mathrm{M}=(x_1^\mathrm{M},\dots, x_{2N}^\mathrm{M})^\mathrm{T} $ of $ -2\mathsf{T}_\mathrm{QI} $ with $ \gamma=0 $ are
    	\begin{align}
    		\text{energy: }&& &E_\mathrm{M}=\pm 2\sqrt{h^2+J^2+2hJ\cos\theta_\mathrm{M}}\\
    		\text{state: }&& &\hspace{-1em}
    		\begin{cases}
    			x_{2i-1}^\mathrm{M}=\dfrac{J}{h}U_{i-2}\qty(\cos\theta_\mathrm{M})+U_{i-1}\qty(\cos\theta_\mathrm{M})\\[2ex]
    			x_{2i}^\mathrm{M}=-\dfrac{E_\mathrm{M}}{2h}U_{i-1}\qty(\cos\theta_\mathrm{M})
    		\end{cases}
    	\end{align}
	    where $ U_j(\cos\theta) $ is the $ j $th Chebyshev polynomial of the second kind and $ y_\mathrm{M}=\cos\theta_\mathrm{M} $ is the largest solution of
    	\begin{equation}
    		hU_{N}\qty(y)+JU_{N-1}\qty(y)=0.
    	\end{equation}
    	Note that $ x_1^\mathrm{M}=-x_{2N}^\mathrm{M}=1 $. Then the perturbed eigenvalue is given by
        \begin{align}
            E_\mathrm{corr}&=E_\mathrm{M}+\dfrac{1}{\vb*{x}^{\mathrm{M}\dagger} \vb*{x}^\mathrm{M}}\vb*{x}^{\mathrm{M}\dagger}\mathrm{diag}(-2\mathrm{i}\gamma,0,\dots, 0,-2\mathrm{i}\gamma)\vb*{x}^\mathrm{M}+\order{\gamma^2}\nonumber\\
            &=E_\mathrm{M}-\dfrac{4\mathrm{i}\gamma}{\vb*{x}^{\mathrm{M}\dagger} \vb*{x}^\mathrm{M}}+\order{\gamma^2}.
        \end{align}
        Note that $ \cos\theta_\mathrm{M}\simeq 1 $ and therefore $ E_\mathrm{M}\simeq 2(h+J) $ for $ N\gg 1 $, and $ U_j(1)=j+1 $. Then,
        \begin{align}
            \vb*{x}^{\mathrm{M}\dagger} \vb*{x}^\mathrm{M}&\simeq\sum_{j=1}^{N}\qty{\qty[\dfrac{J}{h}(j-1)+j]^2+\qty(1+\dfrac{J}{h})^2 j^2}\\
            &=\dfrac{2}{3}\qty(1+\dfrac{J}{h})^2 N^3+\order{N^2}
        \end{align}
        Therefore, $ g $ scales as $ \sim N^{-3} $ for small $ \gamma $. One can see in Fig. \ref{fig:Liouvillian_gap_scaling} that the $ N^{-3} $ scaling of the gap holds well. Extensive numerical calculations show that this is also the case for considerably large $ \gamma $.
        
    \subsection{Plane wave ansatz}\label{subsec:plane_wave}
        When $ h=J $, i.e., the critical case, we can investigate the Liouvillian spectrum in a different way. Since the overall scale is not important for the analysis, we set $ h=J=1 $ in the following. We consider the eigenvalue problem $ \mathsf{T}_\mathrm{QI}\vb*{v}=\lambda\vb*{v} $ which in components reads
        \begin{align}
            v_{j-1}+v_{j+1}=\lambda v_j \quad (j=1,\dots, 2N),\label{eq:eigenvalue_problem_SSH}
        \end{align}
        where we have defined
        \begin{align}
            v_0\coloneqq \mathrm{i}\gamma Q v_1,\quad v_{2N+1}\coloneqq \mathrm{i}\gamma v_{2N}.\label{eq:eigenvalue_problem_boundary_condition}
        \end{align}
        Then, we introduce the following plane wave ansatz
        \begin{align}
            v_j=A z^j+B z^{-j},
        \end{align}
        where $ z\in \mathbb{C} $. From Eq.~(\ref{eq:eigenvalue_problem_SSH}), the eigenvalue can be written as
        \begin{align}
            \lambda=z+z^{-1}.
        \end{align}
        From the boundary condition Eq.~(\ref{eq:eigenvalue_problem_boundary_condition}), one obtains
        \begin{align}
            \begin{cases}
                A+B=\mathrm{i}\gamma Q (Az+Bz^{-1})\\
                Az^{2N+1}+Bz^{-(2N+1)}=\mathrm{i}\gamma (Az^{2N}+Bz^{-2N})
            \end{cases}
        \end{align}
        This equation has a nontrivial solution if the following condition is satisfied:
        \begin{gather}
            \begin{vmatrix}
                 1-\mathrm{i}\gamma Qz & 1-\mathrm{i}\gamma Qz^{-1}\\
                 z^{2N+1}-\mathrm{i}\gamma z^{2N} & z^{-(2N+1)}-\mathrm{i}\gamma z^{-2N} 
            \end{vmatrix}
            =0,
        \end{gather}
        which can be rewritten as
        \begin{align}
            z^{2N}\qty[z-\mathrm{i}\gamma (Q+1)-Q\gamma^2 z^{-1}]-z^{-2N}\qty[z^{-1}-\mathrm{i}\gamma (Q+1)-Q\gamma^2z]=0.
        \end{align}
        For $ Q=+1 $ sector, one obtains
        \begin{align}
            \qty(z^{2N+1}-\mathrm{i}\gamma z^{2N}-\mathrm{i}\gamma z+1)\qty(z^{2N+1}-\mathrm{i}\gamma z^{2N}+\mathrm{i}\gamma z-1)=0.
        \end{align}
        Similarly to the previous subsection, we shall use perturbation theory to analyze the maximum modulus eigenvalue around the small $ \gamma $. In the unperturbed case, $ z $ corresponding to  the largest energy is $ z=\exp(\mathrm{i}\pi/(2N+1))$, which is one of the solutions of
        \begin{align}
            z^{2N+1}-\mathrm{i}\gamma z^{2N}-\mathrm{i}\gamma z+1=0\label{eq:z}
        \end{align}
        with $ \gamma=0 $. Then, we expand $ z $ in terms of $ \gamma $ as
        \begin{align}
            z=\exp(\dfrac{\mathrm{i}\pi}{2N+1})+\gamma z_1+\order{\gamma^2}.
        \end{align}
        Then, it follows from Eq.~(\ref{eq:z}) that
        \begin{align}
        \begin{split}
             &\qty[\exp(\dfrac{\mathrm{i}\pi}{2N+1})+\gamma z_1]^{2N+1}-\mathrm{i}\gamma\qty[\exp(\dfrac{\mathrm{i}\pi}{2N+1})+\gamma z_1]^{2N}\\
            &\hspace{10em}-\mathrm{i}\gamma \qty[\exp(\dfrac{\mathrm{i}\pi}{2N+1})+\gamma z_1]+1=\order{\gamma^2}.
        \end{split}
        \end{align}
        By considering up to the first order in $ \gamma $, we have
        \begin{align}
            z_1=\dfrac{2}{2N+1}\sin(\dfrac{\pi}{2N+1})\exp(\dfrac{\mathrm{i}\pi}{2N+1}).
        \end{align}
        Therefore, the maximum modulus eigenvalue of $ -2\mathsf{T}_\mathrm{QI} $ which is correct up to the first order in $ \gamma $ is given by
        \begin{align}
            E_\mathrm{corr}&=-2(z+z^{-1})\\
            &=-4\cos(\dfrac{\pi}{2N+1})-\dfrac{8\mathrm{i}\gamma}{2N+1}\sin^2\qty(\dfrac{\pi}{2N+1})+\order{\gamma^2}.
        \end{align}
        Recalling that the first decay mode lives in $ Q=+1 $ sector, i.e., the even-filling sector, both the two eigenvalues which have the smallest imaginary part in absolute value (denoted as red dots in Fig.~\ref{fig:SSH_L_100_fig}) should be filled. As a result, the Liouvillian gap $ g $ for large $ N $ is obtained by multiplying by 2:
        \begin{align}
            g\simeq\dfrac{16\pi^2\gamma}{(2N+1)^3}.
        \end{align}
        
\begin{figure}
            \centering
            \includegraphics[width=0.65\linewidth]{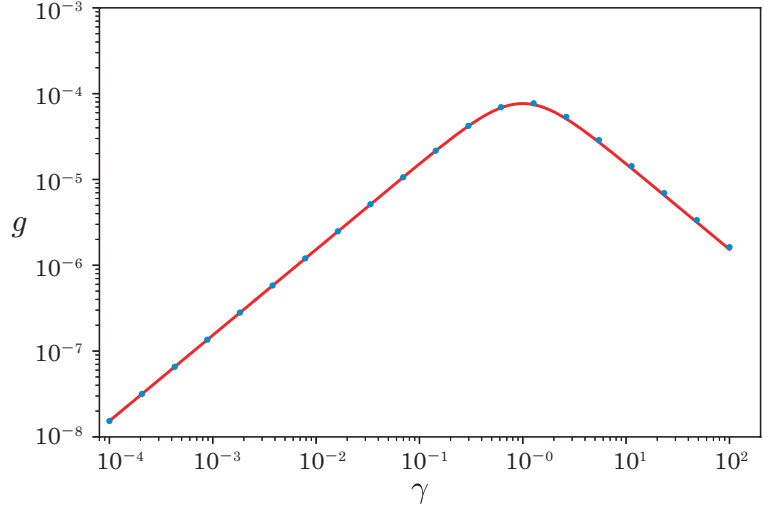}
            \caption{Liouvillian gap as a function of $ \gamma $ for $ h=J=1 $ and $ 2N=100 $. The blue dots are obtained by exact diagonalization, which fit well with the analytical result Eq.~(\ref{eq:Liouvillian_gap_for_finite_gamma}) shown by the red curve.}
            \label{fig:Liouvillian_gap_scaling_vs_gamma_critical}
        \end{figure}

        Moreover, under the condition of $ N \gg 1 $, we can obtain the approximate Liouvillian gap for finite $ \gamma $ following the approach presented in~\cite{Alcaraz2017}. Defining $ k $ as $ z=\mathrm{e}^{\mathrm{i}k} $, Eq.~(\ref{eq:z}) reads
        \begin{align}
            \cos\qty[\qty(N+\dfrac{1}{2})k]=\mathrm{i}\gamma \cos\qty[\qty(N-\dfrac{1}{2})k].
        \end{align}
        One can assume
        \begin{align}
            k=\dfrac{\pi}{2N+1}+\dfrac{\pi\kappa}{2N+1}+\order{N^{-3}},
        \end{align}
        where $ \kappa\sim\order{N^{-1}} $ has been introduced, and obtains the equation for $ \kappa $ by considering up to $ \order{N^{-1}} $ terms:
        \begin{align}
            \cot(\dfrac{\pi\kappa}{2})=\dfrac{\mathrm{i}\gamma+\cos[\pi/(2N+1)]}{\sin[\pi/(2N+1)]}.
        \end{align}
        The solution is
        \begin{align}
        \begin{split}
             \kappa&=\dfrac{1}{2N+1}+\dfrac{2}{\pi}\arctan\qty[\dfrac{\gamma-\mathrm{i}}{\gamma+\mathrm{i}}\tan\dfrac{\pi}{2(2N+1)}]\\
            &=\dfrac{2}{2N+1}\dfrac{\gamma(\gamma-\mathrm{i})}{\gamma^2+1}+\order{N^{-2}}.
        \end{split}
        \end{align}
        and the corresponding eigenvalue of $ -2\mathsf{T}_\mathrm{QI} $ is
        \begin{align}
            E_\mathrm{corr}&=-2\qty{\exp\qty[\dfrac{\mathrm{i}\pi}{2N+1}(1+\kappa)]+\exp\qty[-\dfrac{\mathrm{i}\pi}{2N+1}(1+\kappa)]}\\
            &\simeq -4\cos(\dfrac{\pi}{2N+1})+\dfrac{4\pi\kappa}{2N+1}\sin\dfrac{\pi}{2N+1}\\
            &\simeq -4\cos(\dfrac{\pi}{2N+1})+\dfrac{8\pi^2}{(2N+1)^3}\dfrac{\gamma(\gamma-\mathrm{i})}{\gamma^2+1}.
        \end{align}
        By multiplying the imaginary part by 2 for the same reason as above, we obtain
        \begin{align}
            g\simeq\dfrac{\gamma}{1+\gamma^2}\dfrac{16\pi^2}{(2N+1)^3}.\label{eq:Liouvillian_gap_for_finite_gamma}
        \end{align}
        One can see in Fig.~\ref{fig:Liouvillian_gap_scaling_vs_gamma_critical} that the results show an excellent agreement with numerical results. Interestingly, this result is almost the same as Eq.~(19) in Ref.~\cite{Znidaric2015}, in which the Liouvillian gap of the XX chain with boundary dephasing was studied.
        
\section{Conclusion}\label{sec:conclusion}
    We studied a quantum Ising chain with boundary dephasing as a pedagogical example of exactly solvable dissipative model. The model reduces to a non-Hermitian free-fermion model, the SSH model with imaginary chemical potential at the edges, and it allowed us to investigate the spectrum of Liouvillian in detail. We showed both analytically and numerically that the Liouvillian gap scales with the system size $ N $ as $ N^{-3} $, irrespective of the strength of $ \gamma $.

\section*{Acknowledgment}
We thank Manas Kulkarni, Toma\v{z} Prosen, and Juan P. Garrahan for fruitful discussions. N.S. acknowledges support of the Materials Education program for the future leaders in Research, Industry, and Technology (MERIT).
H. K. was supported in part by JSPS Grant-in-Aid for Scientific Research on Innovative Areas No. JP20H04630, JSPS KAKENHI Grant No. JP18K03445, and the Inamori Foundation.
%Insert the Acknowledgment text here.

\appendix

\section{Quantum compass model with boundary dephasing}\label{app:qunatum_compass}
    The one-particle Hamiltonian which is the same as Eq.~(\ref{eq:SSH_QI}) arises from another dissipative model, namely, one-dimensional quantum compass model with boundary dephasing. A Hamiltonian and Lindblad operators are written as
	\begin{gather}
		H=H_\mathrm{QC}=-\sum_{i=1}^{N/2} J_x\sigma_{2i-1}^x\sigma_{2i}^x -\sum_{i=1}^{N/2-1} J_y\sigma_{2i}^y\sigma_{2i+1}^y,\\
		L_1=\sqrt{\gamma_\mathrm{L}}\sigma_1^z, \quad L_2=\sqrt{\gamma_\mathrm{R}}\sigma_N^z,
	\end{gather}
	where we assume the number of sites $ N $ to be even. Along the same lines as in the authors' previous work on the same Hamiltonian with the bulk dephasing~\cite{Shibata2019}, one can map the Liouvillian to a non-Hermitian Hamiltonian
	\begin{align}
		\mathcal{H}=\mathcal{H}_\mathrm{QC}=2J_x\sum_{i=1}^{N/2}\qty(c_{2i-1}^\dagger c_{2i}+\textrm{h.c.})+2J_y\sum_{i=1}^{N/2-1} \qty(c_{2i}^\dagger c_{2i+1}+\textrm{h.c.})\\
		\hspace{1em}+2\mathrm{i}\mu_1\gamma_\mathrm{L}\qty(c_1^\dagger c_1-\dfrac{1}{2}) +2\mathrm{i}\mu_N\gamma_\mathrm{R}\qty(c_N^\dagger c_N -\dfrac{1}{2})-\mathrm{i}(\gamma_\mathrm{L}+\gamma_\mathrm{R}),
	\end{align}
	where $ \mu_1 $ and $ \mu_N $ can be either $ +1 $ or $ -1 $. As noted in Ref.~\cite{Shibata2019}, the spectrum of $ \mathcal{H}_\mathrm{QC} $ is invariant under $ (\mu_1,\mu_N)\leftrightarrow (-\mu_1, -\mu_N) $, so we now fix $ \mu_N=+1 $. After the charge conjugation $ c_j\to c_j^\dagger $, $ \mathcal{H}_\mathrm{QC} $ in the new basis reads
	\begin{align}
		\mathcal{H}'_\mathrm{QC}=-2\vb{c}^\dagger \mathsf{T}_\mathrm{QC} \vb{c} +\mathrm{i}\gamma_\mathrm{L}(\mu_1-1),\label{eq:H_QC_SSH}
	\end{align}
	where
	\begin{align}
		\mathsf{T}_\mathrm{QC}\coloneqq
		\begin{pmatrix}
			\mathrm{i}\mu_1\gamma_\mathrm{L} & J_x & & & &\\
			J_x & 0 & J_y & & &\\
			& J_y & 0 & \ddots & &\\
			& & \ddots & \ddots & \ddots &\\
			& & & \ddots & 0 & J_x\\
			& & & & J_x & \mathrm{i}\gamma_\mathrm{R}
	\end{pmatrix}.\label{eq:SSH_QC}
	\end{align}
	Comparing Eqs.(\ref{eq:H_QI_SSH}, \ref{eq:SSH_QI}) and Eqs. (\ref{eq:H_QC_SSH}, \ref{eq:SSH_QC}), one can identify $ \mathcal{H}_\mathrm{QI} $ with $ \mathcal{H}_\mathrm{QC} $ via the correspondence $ Q\leftrightarrow \mu_1 $.\footnote{In fact, there is a subtle difference between them. As $ Q $ is a parity operator, after one-particle energies of $ \mathsf{T}_\mathrm{QI} $ are obtained with $ Q $ fixed to $ +1\, (-1) $, then many-body eigenvalues of $ \mathcal{H}_\mathrm{QI} $ should be obtained by filling an even (odd) number of levels. On the other hand, eigenvalues of $ \mathcal{H}_\mathrm{QC} $ are obtained by considering all possible combinations of one-particle energies of $ \mathsf{T}_\mathrm{QC} $ with both $ \mu_1=\pm 1 $.}


\bibliographystyle{ptephy}
\bibliography{ref}
%
% once the .bbl file has been generated then place the text in your article.

\end{document}